\documentclass[12]{article}
\usepackage{amssymb}

\evensidemargin \oddsidemargin
\newcommand{\beq}{\begin{equation}}
\newcommand{\eeq}{\end{equation}}
\newcommand{\ba}{\begin{array}}
\newcommand{\ea}{\end{array}}
\newcommand{\bea}{\begin{eqnarray}}
\newcommand{\eea}{\end{eqnarray}}
\def\b#1{{\mathbb #1}}
\def\nn{\nonumber  \\}
\def\0{{\bf 0}}

\newtheorem{theorem}{Theorem}[section]

\date{}

\begin{document}

\title{Qualitative properties for a class of non-autonomous
semi-linear 3$^{rd}$ order PDE arising in dissipative
problems\footnote{Talk given at the ``15th International Conference on Waves and Stability in Continuous Media'' (WASCOM09), Mondello, Palermo (Sicily), 28/6-1/07/2009. To appear in the Proceedings. }}

 \author{ {\sc  A. D'Anna   \hspace{30mm} G. Fiore}  \\\\
  Dip. di Matematica e Applicazioni, Fac.  di Ingegneria\\
        Universit\`a di Napoli, V. Claudio 21, 80125 Napoli
       }
 \maketitle

\begin{abstract}
We improve results \cite{DanFio09,DacDan98,DanFio00,DanFio05}
regarding the stability and attractivity of solutions $u$
of a large class of initial-boundary-value problems of the form
\bea
&& \left\{\ba{l} -\varepsilon(t)\,
u_{xxt}+u_{tt}-C(t)\,u_{xx}+(a'\!\!+\!a)u_t=F(u),\qquad
 x\!\in\!]0,\!\pi[,\:\: t\!>\!t_0, \\[8pt]
u(0,t)=0, \quad u(\pi,t)=0,  \ea\right.\qquad \label{eq'}  \\[8pt]
&& \quad u(x,\!t_0)\!=\! u_0\!(x), \:\: u_t(x,\!t_0)\!=\! u_1\!(x),\quad\mbox{with }
u_0\!(0)\!=\!u_1\!(0)\!=\!u_0\!(\pi)\!=\!u_1\!(\pi)\!=\!0.\label{eq2'}
 \eea
Here $t_0\!\ge\! 0$,
$\varepsilon\!\in\! C^2\!(I\!,\!I)$, $C\!\in\!C^1\!(I\!,\!\b{R}^+)$
(with $I\!:=\![0,\infty[$) are functions of $t$,
with $C(t)\!\ge\! \overline{C}\!=\!\mbox{const}\!>\!0$;
% the conservative force fulfills
$F(0)\!=\!0$, so that (\ref{eq'}) admits
the null solution $u^0(x,t)\!\equiv\! 0$; $a'\!=\!\mbox{const}\!\ge\!0$,
$a\!=\!a(x\!,\!t\!,\!u\!,\!u_x\!,\!u_t\!,\!u_{xx})\!\ge\!0$, $\varepsilon(t)\!\ge\! 0$.
%, so that the corresponding terms are dissipative.
In the proof we use Liapunov functionals
$W$ depending on two parameters, which we adapt to the `error'
$\sigma$.
\end{abstract}

 \vspace{1mm}


\section{Introduction}\label{aba:sec1}

The class  (\ref{eq'}-\ref{eq2'}) includes (see e.g. the
introduction of \cite{DanFio09}) equations arising in Superconductor
Theory  \cite{Jos,BarPat82,ChrScoSoe99} and in the Theory of
Viscoelastic Materials \cite{Ren83}. We generalize theorem 3.1 of
\cite{DanFio09}, to which we refer also for examples. To formulate
the notions of stability and attractivity\cite{Yos66,FlaRio96} we
use the distance $d(t)\!:=\!d(u,u_t,t)$ between $u,u^0$, where the
norm $d(\varphi,\psi,t)$ is defined by
\beq \ba{l}
d^2(\varphi,\psi,t):=\int_0^\pi[\varepsilon^2(t)
\varphi_{xx}^2\!+\!\varphi_x^2\!+\!\varphi^2\!+\!\psi^2]dx.
\ea\eeq
 $\varepsilon^2$ plays the role of a $t$-dependent weight for
$\varphi_{xx}^2$;
for $\varepsilon\!\equiv\!0$, $d$ reduces to the norm needed
for the corresponding second order problem.
The vanishing of $\varphi,\psi$ in $0,\pi$ implies
$|\varphi(x)|,\varepsilon(t)|\varphi_x(x)|\le d(\varphi,\psi,t)$
%\footnote{From $\varphi^2(x)\!=\!\int_0^xdx'\frac {d\varphi^2}{dx}(x')
%\!=\!\int_0^xdx'2\varphi(x')\varphi_x(x')$ and $2|\varphi\varphi_x|\!\le\!(\varphi
%\!+\!\varphi_x^2)$ it follows $\varphi^2(x)\!\le\!\int_0^xdx'
%(\varphi\!+\!\varphi_x^2)
% \!\le\!\int_0^\pi dx'(\varphi\!+\!\varphi_x^2)\le d^2(\varphi,\psi,t)$;
%similarly one proves $\varepsilon|\varphi_x(x)|\le d(\varphi,\psi)$.}
for all $x$; a convergence w.r.t. $d$ therefore implies a
uniform (in $x$) pointwise convergence of  $\varphi$, and also of
$\varphi_x$  if $\varepsilon(t)\!\neq\! 0$.
%Fix once and for all $\kappa\in\b{R}$, $\xi\!>\!0$.
Throughout the paper
 $t_0\!\in\! I_\kappa\!:=\![\kappa,\infty[$, $\kappa\in\b{R}$, $\xi\!>\!0$.
For any function $f(t)$ we denote $\overline{f}=\inf_{t\!>\!0}f(t)$,
$\overline{\overline{f}}=\sup_{t\!>\!0}f(t)$.

\bigskip\noindent
{\bf Def. 1.1} $u^0$ is {\it stable}  if for any $\sigma\!\in\!]0,\xi]$
there exists a $\delta(\sigma,t_0)\!>\! 0$ such that
\beq
%d(u_0,u_1)\equiv
d(t_0)<\delta(\sigma,t_0)\qquad \qquad\Rightarrow\qquad \qquad
d(t)<\sigma\:\quad\forall t\ge t_0\!\in\! I_\kappa.
\eeq
$u^0$  is {\it uniformly stable}
if $\delta$ can be chosen independent of $t_0$,
$\delta=\delta(\sigma)$.

\medskip\noindent
{\bf Def. 1.2} $u^0$ is {\it %(uniformly)
asymptotically stable} if it is %(uniformly)
stable and
$ \forall t_0\!\in\!I_\kappa$, $\nu\!>\!0$
there exist $\delta(t_0)\!>\!0$, $T(\nu,t_0,u_0,u_1)\!>\! 0$
%($\delta\!>\!0$, $T(\nu)\!>\! 0$):
such that:
\beq
d(t_0)<\delta\qquad \qquad\Rightarrow\qquad \qquad
d(t)<\nu\:\quad\forall t\ge t_0+T.
\eeq
%$u^0$ is {\it uniformly asymptotically stable} if moreover this holds
%with $\delta\!,\!T$ independent of $t_0\!,\! u_0\!,\!u_1$.
%, i.e. $d(t)\to 0$ as $t\to\infty$ uniformly in $t_0,u_0,u_1$.

 \medskip\noindent
{\bf Def. 1.3 } $u^0$
is {\it uniformly exponential-asymptotically stable} if
%there exist positive  constants
$\exists\delta,\! D,\!E\!>\!0$:
%such that
\beq
 d(t_0)<\delta\qquad \Rightarrow\qquad d(t)\leq D
   \exp \left[-E(t-t_0) \right] d(t_0), \quad   \forall
   t\geq t_0\!\in\! I_\kappa.                      %\label{range'}
\eeq

\section{Main assumptions and preliminary estimates}
\label{preliminaries}

\noindent
{\bf Assumptions I:} We assume that there exist constants
$k\!\ge\!0$, $h\!\ge\!0$,  $A\!\ge\!0$, $\omega\!>\!0$,  $\rho\!>\!0$, $\mu\!>\!0$,
 $\tau\!>\!0$ such that
\bea
&& F(0)\!=\!0,\qquad \qquad F_z(z)\le k\!+\!h |z|^\omega\quad\quad\:\:
\mbox{ if } |z|\!<\!\rho.
\label{condi3}\\[8pt]
&& \overline{C}\!>\!k,\qquad
C\!-\!\dot\varepsilon\!\ge\!\mu(1\!+\!\varepsilon),\qquad
\mu\!+\!\overline{C}/2
\!-\!2k\!>\!0,  \qquad \overline{\ddot\varepsilon}\!>\!-\infty.
\label{condi2}\\[8pt]
&& 0\le a\!\le\! A d^\tau(u,u_t,t),\qquad\qquad a'\!+\!
\overline{\varepsilon}/2\!>\!0\label{condi1}
\eea
(we are not excluding $a'\!<\!0$). Setting $h=0$ in (\ref{condi3})
one obtains the analog assumption considered in Ref.\cite{DanFio09}~;
the present one is slightly more general as it may be satisfied with
a smaller $k$, what makes (\ref{condi2})$_1$ weaker,
and/or a larger $\rho$.  Upon integration (\ref{condi3})
implies for all $|\varphi|\!<\!\rho$
\beq
\ba{l}
\varphi F(\varphi)\!\le\!
k\varphi^2+\frac h{\omega\!+\!1} |\varphi|^{\omega\!+\!2},\qquad\quad
\int_0^{\varphi}\!\!F(z)dz\le k\frac {\varphi^2}2+\frac
{h|\varphi|^{\omega\!+\!2}}{(\omega\!+\!1)
(\omega\!+\!2)}.
\ea\label{conseq}
\eeq

We recall Poincar\'e inequality, which easily follows from Fourier analysis:
\beq
\ba{l}\phi\!\in\! C^1\!(]0,\pi[),\:\:\phi(0)\!=\!0,\:\:
\phi(\pi)\!=\!0,\qquad\Rightarrow \qquad\!\! \int\limits^\pi_0 \!
\phi_x^2\!(x)dx\ge \int\limits^\pi_0 \!\phi^2\!(x) dx.
\ea\label{poinc}
\eeq

We introduce the non-autonomous {\bf family of Liapunov functionals}\cite{DanFio09}
$$\ba{l}
W(\varphi,\psi,t;\!\gamma,\theta)\!=\!\!
\displaystyle\int\limits_0^\pi
\!\!\left[\!\gamma\psi^2\!\!+\!(\!\varepsilon
\varphi_{xx}\!\!\!-\!\!\psi\!)^2\!\!\!+\!\!
[C\!(\!1\!\!+\!\!\gamma\!)
%\\[8pt]\qquad\qquad
\!\!+\!\!\varepsilon(\!a'\!\!\!+\!\!\theta\!)\!-\!\dot\varepsilon]\!\varphi_x^2
\!\!+\!\!a'\!\theta\varphi^2\!\!\!+\!\!2\theta\varphi\psi \!-\!
2(\!1\!\!+\!\!\gamma\!)\!\!\displaystyle\int_0^{\varphi(\!x\!)}
\!\!\!\!\!\!\!\!\!F(\!z\!)\!dz\!\right]\!\frac{dx}{2}
\ea%\label{321}
$$
depending on two for the moment  unspecified positive parameters
$\theta,\gamma$. Let $W(t;\gamma,\theta)\!:=\!W(u,u_t,t;\gamma,\theta)$.
In Ref. \cite{DanFio09} we have found
$$\ba{l}
\dot W\!=\!-\!\displaystyle\int\limits_0^\pi\!\! \!\Big\{\!\varepsilon\gamma u^2_{xt}\!+\!\!
\left[\!(\!a\!\!+\!\!a'\!)(\!1\!\!+\!\!\gamma\!)\!\!-\!\!\theta\!\!-\!\!
\frac{\varepsilon a^2}{C\!-\!\dot\varepsilon}\!\!-\!\!\frac{\theta a^2}C
\!\!\right]\!\!u_t^2\!\! +\!\varepsilon(C\!\!-\!\!\dot\varepsilon\!)\!
\!\left[\frac{a u_t}{C\!-\!\dot\varepsilon}\!-\!\frac{u_{xx}}2\!\right]^2\!\!
\!+\!\frac {3\varepsilon}4(C\!-\!\dot\varepsilon)\! u^2_{xx}\\[8pt]
\left.\!\!+\!\!\left[\!C\!\!\left(\!\frac {\theta}2\!\!-\!\!a'\!\right)\!
\!+\!\!\ddot\varepsilon\!+\!(C\!\!-\!\!\dot\varepsilon\!)\!(\!a'\!\!+\!\theta\!)
\!\!-\!\!(\!1\!\!+\!\!\gamma\!)\dot C\!\!-\!\!2\varepsilon
F_u\!\right]\!\!\frac{u_x^2}2\!+\!\frac{\theta C}4(u_x^2\!\!-\!u^2)
\!+\!\frac{\theta C}4\!\left[\!u\!+\!\frac {2a} C u_t\!\right]^2\!\!
\!\!-\!\!\theta uF\!\right\}\!dx
\ea
$$
Provided $|u|\!<\!\rho$, $\theta\!>\!\mbox{max}\{2a'\!,\!-\!a'\}$,
$\mu(a'\!+\!\theta)\!>\!2 k$, (\ref{poinc}) with $\phi\!=\!u_t,u$,
implies
 \bea&&\ba{l}
\dot W \!\le\! -\!\!\displaystyle\int_0^\pi\!
\!\!\left\{\!\!\left[\overline{\varepsilon}\gamma\!+\!(a\!+\!a')(1\!+\!\gamma)
\!-\!\theta\!-\!a^2\!\!\left(\!\!\frac 1{\mu}
\!+\!\frac{\theta}{\overline{C}}\!\!\right)\!\right]\!u_t^2
\!+\!\frac 34 \mu\varepsilon^2 u^2_{xx}\!+\!\bigg[\overline{C}\!\left(\!\frac {\theta}2\!-\!a'\!\!\right)
\!\!+\!\overline{{\ddot\varepsilon}}\right.\\
\left.+\!\mu(a'\!+\!\theta)\!+\!
[\mu(a'\!+\!\theta)\!-\!2 (k\!+\!h|u|^\omega)]\varepsilon
\!-\!(1\!+\!\gamma)\dot C\bigg]\!\frac{u_x^2}2-\theta\!\left(\!k u^2\!+\!\frac h{\omega\!+\!1}
|u|^{\omega\!+\!2}\!\right)\!\!
\right\}\!\!dx \\ \!\!\le\!\! -\!\!\!\displaystyle\int_0^\pi\!
\!\!\left\{\!\!\left[\overline{\varepsilon}\gamma\!+\!(a\!+\!a')(1\!+\!\gamma)
\!-\!\theta\!-\!a^2\!\!\left(\!\!\frac 1{\mu}
\!+\!\frac{\theta}{\overline{C}}\!\!\right)\!\!\right]\!u_t^2
\!+\!\frac {3 \mu}4\varepsilon^2\! u^2_{xx}\!+\!\bigg[\!\!\theta\!\left(\!\!\mu\!+\!
\frac {\overline{C}}2\!-\!2k\!\!\right)\!\!+\!\overline{\ddot\varepsilon}\right.
\ea\nn&&\ba{l}
\left.\!-\!(1\!+\!\gamma)\dot C\!+\!a' (\mu\!-\! \overline{C})\!+\!
[\mu(a'\!+\!\theta)\!-\!2 k]\varepsilon\bigg]\!\frac{u_x^2}2
\!-\!h\varepsilon|u|^\omega u_x^2\!-\!\frac {h\theta}{\omega\!+\!1}
|u|^{\omega\!+\!2}\! \right\}\!dx.\ea\quad \label{ineq0}
\eea
To find an upper bound for $\dot W$ we make {\bf Assumption II}:
\beq
\forall \gamma>0\quad \exists \bar t(\gamma)\!\in\! [0,\infty[\quad
\mbox{such that }
\dot C(1+\gamma)\!\le\!1\quad \mbox{for }t\!\ge\!\bar t.\label{case1}
\eeq
(\ref{case1}) is fulfilled
by $\bar t(\gamma)\!\equiv\! 0$ if $\dot C\!\le\! 0$, %, whereas it is satisfied
by some $\bar t(\gamma)\!\ge\!0$ if
$\dot C\!\stackrel{t\!\to\!\infty}{\longrightarrow}\!0$.
(\ref{case1}) implies $\overline{\ddot
\varepsilon}\!\le\! 0$:  $\overline{\ddot
\varepsilon}\!>\! 0$  would imply $\dot
\varepsilon\ge\overline{\ddot \varepsilon}t\!+\!\dot\varepsilon(0)$,
$\varepsilon\ge\overline{\ddot
\varepsilon}t^2/2\!+\!\dot\varepsilon(0)t\!+\!\varepsilon(0)$
and by (\ref{condi2})$_2$ that $C$ grows at least quadratically
with $t$, against (\ref{case1}).
We choose
\beq
\ba{l}
\theta>\theta_1:=\max\left\{2a',\frac{2k}
{\mu}\!-\!a' ,\frac{5\!-\!\overline{\ddot
\varepsilon}\!-\! a'(\mu\!-\!\overline{C})}
{\mu\!+\!\overline{C}/2\!-\!2k}\right\},\\[8pt]
\gamma>\gamma_1(\sigma):=\frac {1\!+\!\theta\!+\!\overline{\varepsilon}/2}
{a'\!+\!\overline{\varepsilon}}
+\gamma_{32}\sigma^{2\tau}\qquad\quad \gamma_{32}:=\frac
{A^2}{(a'\!+\!\overline{\varepsilon})}
\left(\frac 1{\mu}\!+\!\frac{\theta}{\overline{C}}\right).
\ea\label{thetadef}
\eeq
These definitions respectively imply,  provided $t> \bar t$ and
$d(t)\!\le\!  \sigma\!<\! \rho$,
\beq \ba{l}\theta\!\left(\!\mu\!+\!\overline{C}/2\!-\!2k\!\right)
\!\!+\![\mu(a'\!+\!\theta)\!-\!2 k]\overline{\varepsilon}
\!+\!\overline{\ddot\varepsilon}\!-\!(1\!+\!\gamma)
\dot C\!+\!a' (\mu\!-\! \overline{C})> 4,\\[8pt]
\overline{\varepsilon}\gamma\!+\!(a\!+\!a')(1\!+\!\gamma)\!-\!\theta
\!-\!a^2\!\left(\!\frac 1{\mu}\!+\!\frac{\theta}{\overline{C}}\!\right)\ge
a'\!+\!\frac {a\!+\!a'\!+\!\overline{\varepsilon}}{a'\!+\!\overline{\varepsilon}}
\left[(1\!+\!\theta\!+\!\overline{\varepsilon}/2)\right.
\\ \left. +\!A^2\left(\frac 1{\mu}\!+\!\frac{\theta}{\overline{C}}
\right)\sigma^{2\tau}\right]\!-\!\theta\!-
\!A^2\!\left(\!\frac 1{\mu}\!+\!\frac{\theta}{\overline{C}}\!\right)d^{2\tau}
\ge 1\!+\!a'\!+\!\overline{\varepsilon}/2>1.
\ea\label{ineq2}
\eeq
If $0\!<\!d(t)\!<\! \sigma$
(\ref{ineq0}), (\ref{ineq2}) imply for all
$t\ge \bar t$ the {\bf upper bound for $\dot W$}
\beq\ba{l}
\dot W(u,u_t,t;\gamma,\theta)  \!\le\! -\eta\, d^2(t)+ \displaystyle\int\limits_0^\pi\!
\!h\left[\varepsilon|u|^\omega u_x^2\!+\!\frac {\theta}{\omega\!+\!1}
|u|^{\omega\!+\!2}\right] dx\\[8pt]
\quad\le\!\! \left[-\eta+\!h 2^{\frac\omega 2}\!\left(\!\varepsilon(t)
 \!+\!\frac {\theta}{\omega\!+\!1} \!\right)\!d^{\omega}\!(t)\! \right]\!d^2(t),
\qquad \quad\eta\!:=\!\min\left\{1,\frac 34 \mu\right\}.
\ea\qquad\label{Ineq1}
\eeq

From the definition of $W$ it immediately follows
\[
\ba{l}
W(\varphi,\psi,t;\gamma,\theta) =\displaystyle
\int\limits_0^\pi\!
\frac{1}{2}\!\left\{\!\left(\!\gamma\!-\!\theta^2\!-\!\frac
12\!\right)\psi^2\!+\!\frac{(\varepsilon \varphi_{xx}\!-2\!\psi)^2}4
\!+\!\frac{(\varepsilon \varphi_{xx}\!-\!\psi)^2}2\!+\!
\varepsilon^2\frac{\varphi_{xx}^2}4
\right.\\ \left.
+\![C(1\!+\!\gamma)\!-\!\dot\varepsilon\!+\!\varepsilon
(a'\!+\!\theta)]\varphi_x^2 \!+\!(\theta a'\!-\!1)\varphi^2\!+\!\left[\theta\psi\!+\!\varphi\right]^2\!  \!-\! 2(1\!+\!\gamma)\!\!
\int_0^{\varphi(x)}\!\!\!\!F(z)dz\!\right\}dx.
\ea
\]
Using (\ref{condi2})$_2$, (\ref{conseq}) and (\ref{poinc}) with
$\phi(x)=\varphi(x)$ we find for $|\varphi|\!<\!\rho$
$$
\ba{l}
 W \!\ge\! \displaystyle\int\limits_0^\pi\!
\!\frac {dx}2\!\!\left\{\!\!\left[\!\gamma\!\!-\!\!\theta^2\!\!\!-\!\!\frac 12\!\right]\!
\!\psi^2\!\!+\!\frac{\varepsilon^2\!\varphi_{xx}^2}4\!+\!\!\left[\!
\mu\!\!+\!\!\!\left(\!\mu\!\!+\!\!a'\!\!+\!\!\frac{\theta}{2}\!\right)\!
\overline{\varepsilon}\!\right]\!\!\varphi_x^2
%\right.\\ \left.
\!\!+\!\!\left[\!\left(\!a'\!\!\!+\!\!\frac{\overline{\varepsilon}}{2}\!\right)
\!\!\theta\!\!-\!\!1\!\!-\!\!k\!\!+\!\!(\overline{C}\!\!-\!\!k)\!\gamma
\!\!-\!\!\frac {2h(1\!+\!\gamma)
|\varphi|^{\omega}}{(\omega\!+\!1)(\omega\!+\!2)}\!\right]\!\!
\varphi^2\!\!\right\}.
\ea
$$
Choosing $\theta\!>\!\theta_2\!:=\!\max\left\{\theta_1,
\frac{\overline{C}\!+\!5/4}{a'\!+\!\overline{\varepsilon}/2}\right\}$,
$\gamma\!\ge\!\gamma_2(\sigma)\!:=\!\gamma_1(\sigma)\!+\!\theta^2\!+\!1$
we find
$$\ba{l}
W\!>\! \displaystyle\int\limits_0^\pi\!
\frac{1}{2}\!\left\{\!\left[\gamma\!-\!\theta^2\!-\!\frac 12\right]
\psi^2\!+\!\varepsilon^2\frac{\varphi_{xx}^2}4\!+\!\left[
\mu\!+\!\!\left(\mu\!+\!a'\!+\!\frac{\theta}{2}\right)\overline{\varepsilon}\right]
\varphi_x^2\right.\\ \qquad\qquad\left.
+\!\left[\frac 14\!+\!(1\!+\!\gamma)\!\left(\overline{C}\!-\!k\!-\!\frac {2h|\varphi|^{\omega}}{(\omega\!+\!1)(\omega\!+\!2)}\!\right)\!\right] \varphi^2\right\}dx.
\ea
$$
By the inequality $|\varphi|\!<\!d$ the expression in the last bracket is positive if
$$
d(t)\le \sigma <\rho_2:=\mbox{min}\left\{\rho,\left[(\overline{C}\!-\!k)(\omega
\!+\!1)(\omega\!+\!2)/2h\right]^{1/\omega}\right\}.
$$
Hence for
$d\!\le\! \sigma$ %the expression in
the last square bracket
is larger than $1/4$, and we find the {\bf lower bound for $W$}
\beq
\ba{l}
W(\varphi,\!\psi,\!t;\!\gamma,\!\theta) \ge \chi
d^2(\varphi,\!\psi,\!t),\qquad \chi\!:=\!\frac 12\min\!\left\{\!\frac
14,\mu\!+\!\!\left(\!\mu\!+\!a'\!+\!\frac{\theta}{2}\right)\!\overline{\varepsilon}
 \!\right\}\!>\!0. \label{Ineq2}
\ea
\eeq

We also recall the {\bf upper bound for $W$} proved in
\cite{DanFio09} for $d\le \sigma$:
\beq
W(\varphi,\psi,t;\gamma,\theta) \le \left[1\!+\!\gamma(\sigma)\right]g(t) B^2(d).
\label{Ineq3}
\eeq
The map $d\!\in\![0,\infty[\to
B(d)\!\in\![0,\infty[$ is  continuous and increasing, hence invertible.
Moreover, $B(d)\ge d$. Here we have chosen $\gamma$ and defined
\beq\ba{l}
\gamma\ge\gamma_3(\sigma):=\gamma_2(\sigma)\!+\!1\!+\!
\frac{a'\!+\!\theta}{\mu}\!+\!(a'\!+\!1)\theta
=\gamma_{31}+\gamma_{32}\sigma^{2\tau},\\[10pt]
\gamma_{31}:=\frac {1\!+\!\theta}{a'\!+\!\overline{\varepsilon}}
\!+\!\theta^2\!+\!2\!+\!
\frac{a'\!+\!\theta}{\mu}\!+\!(a'\!+\!1)\theta, \quad\qquad
g(t)\!:=\!C(t)\!-\!\frac{\dot\varepsilon(t)}2\!+\! 1\!>\!1,\\[10pt]
m(r)\!:=\! \max \!\left\{|F_\zeta(\zeta)| \: : \:  |\zeta |\le r\right\},
\quad\qquad B^2(d):=\left[1\!+\! m(d)\right]d^2.  \ea\label{defgB}
\eeq

Fixed $\sigma\!\in\!]0,\rho_2[$, if $d\!<\!\sigma$
we find $B^2\!(d)\!\le\! [1\!+\! m(\sigma)]d^2$ and, by
(\ref{Ineq1}-\ref{Ineq3}),
\beq
\ba{l}
\dot W< - l W+n W^{1\!+\!\frac{\omega}2},\\[8pt]
n(t) \!:=\!\frac{h 2^{\frac{\omega}2}\!}{\chi^{1\!+\!
\frac{\omega}2}}\!\left[\frac {\theta}{\omega\!+\!1}\!+\!
\varepsilon(t)\right] ,\quad \:
l(t,\sigma)\!:=\!\frac{\lambda(\sigma)}{g(t)},\quad \:
\lambda(\sigma)\!:=\!\frac{\eta} {[1\!+\!
m(\sigma)][1\!+\!\gamma_3(\sigma)]}. \ea \label{Ineq4}
\eeq
$\lambda(\sigma)$ is positive-definite and decreasing. By the
Comparison Principle \cite{Yos66}, $W(t)\!<\! y(t)$ for $t>t_0$,
where $y(t)$ solves the Cauchy problem
$$
\dot y= -l y\!+\!n y^{1\!+\!\omega/2}, \qquad \qquad y(t_0)=W_0:=W(t_0)
$$
and we have to choose $t_0\ge \bar t$.
As known, the change of variable $z=y^{-\omega/2}$ reduces this
Bernoulli equation  to the linear one
$\dot z= zl\omega/2 \!-\!  n\omega/2$, which is easily solved to
give the following {\bf comparison equation for  $W$}
for $t\!>\!t_0$:
\beq
\ba{l}
W(t)\!<\! y(t)=W_0\, e^{-\!\lambda\!\int\limits^t_{t_0}\!\!\!\frac{d\tau}{g(\tau)}}
\!\left\{\!1\!-\!W_0^{\frac{\omega}2}\frac{\omega}2\!\!\int\limits^t_{t_0}\!\!n(\tau)
e^{-\!\frac{\omega\lambda} 2\!\! \int^{\tau}_{t_0}\!\frac{d\tau'}{g(\tau')}}\!
d\tau\!\right\}^{-\frac 2{\omega}}
\ea\label{ComparPrinc}
\eeq
%From (\ref{Ineq4}), $\dot W(t)$ is negative (non-positive) if
%$l(t)/n(t)>[W(t)]^{\frac{\omega}2}>0$; therefore   in Case 2
% we have to require   $\dot C_+(t)<2\eta/(1\!+\!\gamma)$.
A sufficient condition for  $\dot W(t)$ to be negative is  that
$n/l<W^{-\frac{\omega}2}$, namely
$$
\frac{n(t)g(t)}{\lambda}<W_0^{-\frac{\omega}2}e^{\frac{\omega\lambda}2\int^t_{t_0}\!\!
\frac{d\tau}{g(\tau)}}
\left\{\!1\!-\!W_0^{\frac{\omega}2}\frac{\omega}2\!\int^t_{t_0}\!\!n(\tau)
e^{-\frac{\omega\lambda} 2 \!\int^{\tau}_{t_0}\!\frac{d\tau'}{g(\tau')} }
d\tau\!\right\},
$$
or equivalently, after some algebra, that
\beq
\ba{l}
W_0^{-\frac {\omega}2}>s(t;t_0,\sigma),\\[10pt]
s(t;t_0,\sigma)\!:=\!\frac{n(t)g(t)}{\lambda(\sigma)}
e^{-\frac{\omega\lambda(\sigma)}2\int^t_{t_0}\!\frac{d\tau}{g(\tau)}}
\!+\!\frac{\omega}2\int^t_{t_0}\!\!n(\tau)e^{-\frac{\omega\lambda(\sigma)}
2 \int^{\tau}_{t_0}\!\frac{d\tau'}{g(\tau')}}d\tau.
\ea \label{Ineq4'}
\eeq
Summing up, $W(t)$ is decreasing and fulfills (\ref{ComparPrinc})  in $[t_0,\infty[$
if $d(t)<\sigma$ and (\ref{Ineq4'}) is satisfied for all $t\ge t_0$, or equivalently if
\beq
S(t_0,\sigma)\!:=\!\sup_{[t_0,\infty[}s(t;t_0,\sigma)\!<\!\infty, \qquad
 \Delta(t_0,\sigma)\!:=\! S(t_0,\sigma)W_0^{\frac {\omega}2}\!<\!1.  \label{Ineq5}
\eeq
We give upper bounds for $s(\!t\!;\!t_0\!,\!\sigma\!)$,
$S(\!t_0\!,\!\sigma\!)$ using
$g$ only: (\ref{defgB})$_3$, (\ref{condi2})$_2$ imply
$$
\ba{l}
g\!=\!\frac 12[C\!-\!\dot\varepsilon]\!+\! \frac C 2\!+\!1\!\ge\!
\frac {\mu}2(1\!+\!\varepsilon)\!+\! \frac C 2\!+\!1
\qquad\Rightarrow \qquad  0\le n(t)\le \alpha_1[\alpha_2+g(t)],
\ea
$$
where $\alpha_1=\frac{h 2^{1\!+\!\frac{\omega}2}\!}{\mu\chi^{1\!+\!
\frac{\omega}2}}$,
$\alpha_2=\left[\frac {\mu\theta}{\omega\!+\!1}\!-\!\mu\!-\!2\!-\!
\overline{C}\right]/2$. Hence, as announced,
\beq
\ba{l}
s(t;\!t_0,\!\sigma)\!\le\! \frac{\alpha_1}{\lambda}[\alpha_2\!+\!g(t)]g(t)
e^{-\!\frac{\omega\lambda}2\!\!\!\!\int\limits^t_{t_0}\!\!\!\frac{d\tau}{g(\tau)}}
\!\!+\!\frac{\omega}2\!\!\int\limits^t_{t_0}\!\!\alpha_1[\alpha_2\!+\!g(\tau)]
e^{-\!\frac{\omega\lambda}
2 \!\!\int\limits^{\tau}_{t_0}\!\!\!\frac{d\tau'}{g(\tau')}}\!d\tau\\[8pt]
=\frac{\alpha_1}{\lambda}[\alpha_2+g(t_0)]g(t_0)
+\frac{\alpha_1}{\lambda}\int^t_{t_0}\!\!
e^{-\frac{\omega\lambda}2 \int^{\tau}_{t_0}\!\frac{d\tau'}{g(\tau')}}
\dot g(\tau)[\alpha_2+2g(\tau)] d\tau \\%[8pt]
\le\!\frac{\alpha_1}{\lambda}[\alpha_2\!+\!g(t_0)]g(t_0)
\!+\!\frac{\alpha_1}{\lambda}\left[\frac 1{1\!+\!\gamma_3(\sigma)}
\!-\!\frac{\overline{\ddot\varepsilon}}2\right]\!\int^t_{t_0}\!\!\!\!
e^{-\!\frac{\omega\lambda}
2 \int^{\tau}_{t_0}\!\frac{d\tau'}{g(\tau')}}[\alpha_2\!+\!2g(\tau)]
d\tau
\ea\label{temp}
\eeq
where we have integrated by parts and used (\ref{case1}) to get
$\dot g\!=\!\dot C\!-\!\ddot\varepsilon/2\!\le\! 1/(1\!+\!\gamma_3)\!-\!
\overline{\ddot\varepsilon}/2$. As $\overline{\ddot\varepsilon}\!\le\! 0$,
the second square bracket is positive; the last integral is an increasing
function of $t$ as its argument is positive, whence
$$
\ba{l}
S(t_0,\sigma)\!\le\!\frac{\alpha_1}{\lambda}[\alpha_2\!+\!g(t_0)]g(t_0)
+\frac{\alpha_1}{\lambda}\!\left[\!\frac 1{1\!+\!\gamma_3(\sigma)}\!-\!\frac{\overline{\ddot\varepsilon}}2\right]\!\!
\displaystyle\int\limits^{\infty}_{t_0}\!\!
e^{-\!\frac{\omega\lambda}
2 \!\!\int^{\tau}_{t_0}\!\!\frac{d\tau'}{g(\tau')}}[\alpha_2\!+\!2g(\tau)]
d\tau,
\ea$$
and $S(t_0,\sigma)<\infty$ for all $t_0\ge 0$ if
\beq
\ba{l}
G(\sigma):=h\int^{\infty}_0\!\! e^{-\frac{\omega\lambda(\sigma)}
2 \int^{\tau}_0\!\frac{d\tau'}{g(\tau')}}g(\tau)
d\tau<\infty .
\ea\label{Ineq5'}
\eeq
Let $\sigma'_{\scriptscriptstyle M}\!:=\!
\sup\{\sigma\!\in\!\b{R}^+ |\: G(\sigma)\!<\!\infty\}$.
If $h=0$, then $G(\sigma)\!\equiv\! 0$,
$\sigma'_{\scriptscriptstyle M}\!=\!\infty$
and any $W_0$ fulfills (\ref{Ineq5})$_2$. It is
$\sigma'_{\scriptscriptstyle M}\!=\!\infty$ also if $h\!>\!0$ and e.g.
$g(t)\!\le\! K'\!+\!K'' t^a$ with some $K',K''\!>\!0$,
$0\!\le\! a<1$; whereas $h\!>\!0$ and e.g.
$g(t)\!\le\! K'\!+\!K t$ with some $K'\!>\!0$,
$K\!\in\!]0,\!\frac{\omega\lambda(\sigma)}4[$ gives a finite
$\sigma'_{\scriptscriptstyle M}\!>\!0$, determined by
$\lambda(\sigma'_M)=4K/\omega$.

\noindent
The inequality $\sigma'_{\scriptscriptstyle M}\!>\!0$ and (\ref{Ineq5'})
imply $\int^\infty_0 \!\!\frac{dt}{g(t)}\!=\!\infty$: in fact, if it were
$\int^\infty_0 \!\!\frac{dt}{g(t)}\!<\!\infty$ it would be
$e^{-\frac{\omega\lambda(\sigma)}
2 \!\int^{\tau}_0\!\!\frac{d\tau'}{g(\tau')}}>L\!:=\!e^{-\frac{\omega\lambda(\sigma)}
2 \int^{\infty}_0\!\!\frac{d\tau'}{g(\tau')}}\!>\!0$, whence
$G(\sigma)>hL\int^{\infty}_0\!\! g(\tau)d\tau=\infty$, for
{\it all} $\sigma\!>\!0$.

\section{Stability and asymptotic stability of the null solution $u^0$}
\label{stability}

\begin{theorem}
Assume  conditions (\ref{condi3}-\ref{condi1})
and either $\dot C\!\le\! 0$ for all $t\!\in\! I$, or
$\dot C\!\stackrel{t\!\to\!\infty}{\longrightarrow}\!0$. %, are fulfilled.
$u^0$ is stable if
$\sigma'_{\scriptscriptstyle M}\!>\!0$, asymptotically stable if
moreover
$\int^\infty_0 \!\!\frac{dt}{g(t)}\!=\!\infty$.
$u^0$ is uniformly stable and exponential-asymptotically
stable if $\overline{\overline g}<\infty$. \label{thm1}
\end{theorem}

{\bf Proof.} We first analyze the behaviour of
$r^2(\sigma)\!:=\!\frac{\sigma^2}{1\!+\!\gamma_3(\sigma)}
\!=\!\frac{\sigma^2}{1\!+\!\gamma_{31}\!+\!\gamma_{32}\sigma^{2\tau}}$.
By (\ref{defgB})$_1$ the positive constants $\gamma_{31},\gamma_{32}$
are independent of $\sigma, t_0$.
$r(\sigma)$ is an increasing and therefore invertible map
$r\!:\![0,\sigma_{\scriptscriptstyle M}[\to [0,r_{\scriptscriptstyle M}[$, where:
$$ \ba{lll}
\sigma_{\scriptscriptstyle M}\!=\!\infty,\qquad \qquad & r_{\scriptscriptstyle M}\!=\!\infty,\qquad\qquad &
\mbox{if }\:\tau\!\in\![0,1[,\\[8pt]
\sigma_{\scriptscriptstyle M}\!=\!\infty\qquad \qquad & r_{\scriptscriptstyle M}\!=\!1/\sqrt{\gamma_{32}},\qquad\qquad
& \mbox{if }\:\tau\!=\!1,\\[8pt]
\sigma_{\scriptscriptstyle M}^{2\tau}:=\frac{1\!+\!\gamma_{31}}{\gamma_{32}(\tau\!-\!1)},
\qquad \qquad &
r_{\scriptscriptstyle M}\!=\![\frac{\tau\!-\!1}{1\!+\!\gamma_{31}}]^{\frac{\tau\!-\!1}{2\tau}}
/\sqrt{\tau}\gamma_{32}^{\frac
1{2\tau}},\qquad\qquad & \mbox{if }\: \tau\!>\!1,\ea  %\label{defsigmaM}
$$
[in the latter case $r(\sigma)$ is decreasing beyond $\sigma_{\scriptscriptstyle M}$].
Next, let $\xi\!:=\!\min\!\left\{\rho,\sigma_{\scriptscriptstyle M},\sigma'_{\scriptscriptstyle M}\right\}$
if the rhs is finite,
otherwise choose $\xi\in\b{R}^+$; we shall consider an ``error''
$\sigma\!\in]0,\xi[$. We define $\kappa:=\bar t[\gamma_3(\xi)]$ and
\beq
\ba{l}
\delta(\sigma,t_0):=\min\!\left\{B^{-1}\!\!\left[
\frac {\sigma\sqrt{\chi}}{\sqrt{g(t_0)(1\!+\!\gamma_3(\sigma))}}\!
\right]\!, B^{-1}\!\!\left[\frac {[S(t_0,\sigma)]^{-\frac 1{\omega}}}
{\sqrt{g(t_0)(1\!+\!\gamma_3(\sigma))}}\!\right]\!
 \right\}.
\ea\label{defdelta}
\eeq
$\delta(\sigma,t_0)$ belongs to $]0,\sigma[$,
because $d\!\le\! B(d)$ implies $B^{-1}(d)\!\le\! d$, whence
$B^{-1}\!\left[\sigma\sqrt{\chi}/\sqrt{g(t_0)(1\!+\!\gamma_3)}\right]\!\le\!
\sigma/4$, and is an increasing function of
$\sigma$. $\bar t(\gamma)$ was defined in
(\ref{case1}); it is $\bar t[\!\gamma_3\!(\!\sigma\!)\!]\!\le\!\kappa$, as the
function $\bar t[\!\gamma_3\!(\!\sigma\!)\!]$ is non-decreasing. Mimicking an
argument of \cite{DanFio05,DanFio09} we show that for any $t_0\ge\kappa$,
 $\sigma\!\in]0,\xi[$
\beq
d(t_0)<\delta(\sigma,t_0)\qquad \qquad \Rightarrow
\qquad \qquad d(t) <\sigma \qquad\forall t\ge t_0.  \label{subtesi}
\eeq
%Therefore, if $\tau\!>\!1$, $\rho\!>\!\sigma_{\scriptscriptstyle M}$,
%$\sigma\!\in\![\sigma_{\scriptscriptstyle M},\rho[$ one has actually to define
%$\delta(\sigma,t_0)\!:=\!\delta(\sigma_{\scriptscriptstyle M},t_0)$ in (\ref{subtesi}).
{\it Ad absurdum}, assume (\ref{subtesi}) is fulfilled for all
$t\!\in\![t_0,t_1[$ whereas $d(t_1)\!=\!\sigma$,
with some $t_1\!>\!t_0$.
 (\ref{Ineq5}) is trivially satisfied if $h\!=\!0$; if $h\!>\!0$
it follows from
$$
W_0\!\le\![1\!+\!\gamma_3]g(t_0)
B^2\big[d(t_0)\big]<\left[1\!+\!\gamma_3(\sigma)\right]g(t_0)
B^2\big[\delta(\sigma,t_0)\big]\le [S(t_0,\sigma)]^{-\frac 2{\omega}},
$$
where we have used (\ref{Ineq3}), (\ref{defdelta}) in the first and
last inequality. It implies
that $W(t)\equiv W[u,u_t,t;\gamma_3(\sigma),\theta]$ is a decreasing
function of $t$ in $[t_0,t_1]$. Using (\ref{Ineq2}) and again (\ref{Ineq3}),
(\ref{defdelta}) we find the following contradiction with $d(t_1)=\sigma$:
$$
\chi d^2\!(t_1)\le W(t_1)< W_0
<\left[1\!+\!\gamma_3(\sigma)\right]g(t_0)
B^2\big[\delta(\sigma,t_0)\big]\le\chi\sigma^2.
$$
(\ref{subtesi}) amounts to the stability of $u^0$;
if $\overline{\overline g}\!<\!\infty$ we can replace $g(t_0)$ by
$\overline{\overline g}$ in the first inequality of (\ref{temp})
and obtain by integration the stronger inequalities
\beq
\ba{l}
s(t;t_0,\sigma)\le \frac{\alpha_1}{\lambda(\sigma)}
\left[\alpha_2+\overline{\overline g}\right]\,\overline{\overline g}
\qquad\Rightarrow \qquad S(t_0,\sigma)\le \frac{\alpha_1}{\lambda(\sigma)}
\left[\alpha_2+\overline{\overline g}\right]\,\overline{\overline g};
   \label{Ineq7}\ea
\eeq
because of (\ref{Ineq7}) we find the uniform stability (Def. 1.1) with
$$
\ba{l}
\delta(\sigma):=\min\bigg\{B^{-1}\!\left[
\frac {\sigma\sqrt{\chi}}{\sqrt{\overline{\overline g}(1\!+\!\gamma_3(\sigma))}}
\right],\, B^{-1}\!\bigg[\frac {\left[\frac{\alpha_1\overline{\overline g}}{\lambda(\sigma)}\left(
\alpha_2+\overline{\overline g}\right)\right]^{-\frac 1{\omega}}}
{\sqrt{\overline{\overline g}(1\!+\!\gamma_3(\sigma))}}\bigg]
\bigg\}. %, \qquad \qquad \kappa:=\bar t[\gamma_3(\xi)].
\ea
$$

Let now $\delta(t_0)\!:=\!\delta(\xi/2,t_0)$. By (\ref{subtesi})
we find that, for any $t_0\!\ge\!\kappa$,
$d(t_0)\!<\!\delta(t_0)$ implies $d(t)\!<\!\xi/2$ for all $t\!\ge\! t_0$.
Choosing $W(t)\!\equiv\! W[u,\!u_t,\!t;\!\gamma_3(\xi/2),\!\theta]$,
on one hand (\ref{Ineq3}) becomes
$W(t)\!\le\! \frac{\eta g(t)}{\lambda(\xi/2)}d^2(t)$,
while by (\ref{Ineq4'}), (\ref{Ineq5})
$$
\ba{l}
W_0^{\frac {\omega}2}\!s\!\!\left(\!t;\!t_0\!,\!\!\frac{\xi}2\!\right)\!=\!
W_0^{\frac {\omega}2}\!\!\!\left[\frac{n(t)g(t)}{\lambda(\xi/2)}
e^{-\!\frac{\omega}2\!\lambda\left(\!\frac{\xi}2\!\right)\!\!\int^t_{t_0}\!\!\frac{d\tau}{g(\tau)}}\!
+\!\frac{\omega}2\!\!\displaystyle\int^t_{t_0}\!\!\!\!n(\tau)e^{-\!\frac{\omega}2\lambda\left(\!\frac{\xi}2\!\right)\!\! \int^{\tau}_{t_0}\!\!\!\frac{d\tau'}{g(\tau')}}d\tau\right]
\!\le\! \Delta\!\!\left(\!t_0\!,\!\frac{\xi}2\!\right)
\ea
$$
with $\Delta(t_0,\!\xi/2)\!<\!1$, and
$1\!-\!W_0^{\frac {\omega}2}\!\frac{\omega}2\!\!\!\displaystyle\int^t_{t_0}
\!\!\!\!n(\tau)e^{-\!\frac{\omega}2\lambda\left(\frac{\xi}2\right)\!\!
 \int^{\tau}_{t_0}\!\!\frac{d\tau'}{g(\tau')}}d\tau\!\ge\! 1\!-\!
\Delta(t_0,\!\xi/2)\!>\!0$.
These inequalities and (\ref{Ineq2}), (\ref{ComparPrinc})  imply
$$\ba{l}
d^2\!(t)\!\le\! \frac{W(t)}{\chi}\!<\!
 \frac{W_0}{\chi} e^{-\!\lambda\!\!\int\limits^t_{t_0}\!\!\!
 \frac{d\tau}{g(\tau)}}\!\!\left[\!1\!\!-\!\frac{\omega}2W_0^{\frac{\omega}2}
\!\!\!\!\displaystyle\int^t_{t_0}\!\!\!\!\!n(\tau)
e^{\!-\!\frac{\omega\lambda} 2\!\! \!\int\limits^{\tau}_{t_0}\!\!\!\!
\frac{dz}{g(z)}}\!\!d\tau\!\right]^{\!\!-\!\frac 2{\omega}}
\!\!\!\!\!\!\!\!\!<\!\frac{\eta g(\!t_0\!)d^2\!(\!t_0\!)}{\lambda\chi}
\!e^{\!-\!\lambda\!\!\int\limits^t_{t_0}\!\!\! \frac{d\tau}{g(\tau)}}\!\!
\left[\!1\!-\!\Delta\!\!\left(\!t_0,\!\frac{\xi}2\!\right)
\!\right]^{\!-\!\frac 2{\omega}}
\ea
$$
with $\lambda\!=\!\lambda(\xi/2)$. The condition
$\int^\infty_0 \!\!\frac{dt}{g(t)}\!=\!\infty$ implies that the
exponential goes to zero as $t\to\infty$, proving the asymptotic
stability of $u^0$; if $\overline{\overline
g}\!<\!\infty$ we can replace $g(t_0), g(\tau)$ by $\overline{\overline
g}$ in the last inequality and obtain
$$
\ba{l}
d^2(t)< d^2(t_0)\frac{\eta \overline{\overline g}}
{\lambda(\xi/2)\chi}
\exp\left[-\frac{\lambda(\xi/2)}{\overline{\overline g}}(t\!-\!t_0)\right]
\left[\!-\!\Delta(t_0,\xi/2)\right]^{-\!\frac 2{\omega}},
\ea
$$
proving the uniform exponential-asymptotic stability of $u^0$:
set in Def. 1.3
$$
\ba{l}
\delta\!=\!\delta\left(\xi/2,t_0\right),\qquad
D\!=\!\sqrt{\frac{\eta \overline{\overline g}}{\lambda(\xi72)\chi}}\left[1\!-\!\Delta\left(t_0,\frac{\xi}2\right)\right]^{\!-\!\frac 2{\omega}},\qquad
E\!=\!\frac{\lambda(\xi/2)}{2\overline{\overline g}}.
\ea
$$

\end{document}